\documentclass[10pt]{article}
\usepackage{graphicx}
\usepackage{subfigure}
\usepackage{lineno}

\def\Title#1{\begin{center} {\Large #1 } \end{center}}
\def\Author#1{\begin{center}{ \sc #1} \end{center}}
\def\Address#1{\begin{center}{ \it #1} \end{center}}

\newcommand\pubblock{\rightline{\begin{tabular}{l} Proceedings of the Fifth Annual LHCP\\ \pubnumber\\
         \pubdate  \end{tabular}}}

\newenvironment{Abstract}{\begin{quotation} \begin{center} 
             \large ABSTRACT \end{center}\bigskip 
      \begin{center}\begin{large}}{\end{large}\end{center} \end{quotation}}

\newenvironment{Presented}{\begin{quotation} \begin{center} 
             PRESENTED AT\end{center}\bigskip 
      \begin{center}\begin{large}}{\end{large}\end{center} \end{quotation}}

\def\beq{\begin{equation}}
\def\eeq#1{\label{#1}\end{equation}}
\def\eeqn{\end{equation}}

\def\beqa{\begin{eqnarray}}
\def\eeqa#1{\label{#1}\end{eqnarray}}
\def\eeqan{\end{eqnarray}}

\let\bar=\overbar

\def\Dslash{\not{\hbox{\kern-4pt $D$}}}
\def\dslash{\not{\hbox{\kern-2pt $\del$}}}

\def\msb{{\bar{\ssstyle M \kern -1pt S}}}

\usepackage{color}

\newcommand\pubnumber{ ATL-PHYS-PROC-2017-132 }

\newcommand\pubdate{\today}

\def\affiliation{
On behalf of the ATLAS and CMS Collaborations, \\
I.N.F.N. - Sezione di Roma 1 \\
p.le A. Moro 5, Roma, 00185, Italy}

\begin{document}

\large
\begin{titlepage}
\pubblock

\vfill
\Title{  Search for new physics in dijet final states in ATLAS and CMS }
\vfill

\Author{ MATTEO BAUCE  }
\Address{\affiliation}
\vfill
\begin{Abstract}

Events containing a pair of high energy hadronic jet can provide clear signatures in the search for new physics at high energy hadron colliders.
The ATLAS and CMS experiments collected the data from LHC collisions at $\sqrt{s}$= 13 TeV during 2015 and 2016, looking for evidence of new resonances or deviations from the Standard Model predictions. Althoug no hint of new physics was seen, strong limits have been set on the most interesting benchmark models, improving LHC Run1 reach.

\end{Abstract}
\vfill

\begin{Presented}
The Fifth Annual Conference\\
 on Large Hadron Collider Physics \\
Shanghai Jiao Tong University, Shanghai, China\\ 
May 15-20, 2017
\end{Presented}
\vfill
\end{titlepage}
\def\thefootnote{\fnsymbol{footnote}}
\setcounter{footnote}{0}
%

\normalsize 


\section{Introduction}

New and unknown physics has been often revealed by the presence of a new resonant particle, becoming accessible in increasingly higher energy particle collisions. The $pp$ collisions produced by the LHC at $\sqrt{s}$= 13 TeV can therefore contain such evidence of new physics, which is why the ATLAS \cite{Atlas:exp} and CMS \cite{Cms:exp} experiment are scrutinizing them seeking for exotics signals. 
For a new particle produced in hadronic collisions, one of the possible decay mode is into quarks and gluons, making events with a pair of high-energy hadronic jets a good signature for this kind of search. Popular Standard Model (SM) extensions predict a possible signal appearing either as a resonant peak in the dijet invariant mass spectrum or through a peculiar shape of the two jets angular correlation. Here we briefly review the most recent results reported by the ATLAS and CMS experiments in the data collected in the years 2015 and 2016, corresponding to a maximum of $\sim$40 fb$^{-1}$, in several dijet related signatures.

\section{Dijet resonance searches}
\label{sec:resonant}
A new resonant particle decaying to a pair of quarks ($q$) or gluons ($g$) should result in a pair of hadronic jets reconstructed in the detector calorimeter, for which the invariant mass ($m_{jj}$) is expected to peak approximately at the value of the particle mass itself. This peculiar signature has to be distinguished from an overwhelming background of dijet events produced in $pp$ collisions according to Quantum Chromo Dynamics (QCD), which instead has a smoothly falling $m_{jj}$ spectrum. This kind of search can aim at detecting a signal from $q^*$, $W^\prime$, $Z^\prime$, $W^*$, Quantum Black Holes, or Randall-Sundrum grativon benchmark models. 
Similar strategies have been used by ATLAS \cite{atlas:dijet} and CMS \cite{cms:dijetRes} collaborations in such searches, although some different background prediction techniques have been recently introduced. 

CMS selects events with a pair of hadronic jets ($p_T>$30 GeV, $|\eta_j|<$2.5) reconstructed through the particle flow algorithm with a jet-jet rapidity difference satisfying $|\eta_{j_1}-\eta_{j_2}|<$1.3. The online selection of such events is done using a trigger requiring $H_T\equiv\sum_{jets}p_T>$900 GeV, allowing exploration of the kinematic region with $m_{jj}>$1.25 TeV. Close-by jets are merged in single $\Delta R$=1.1 wide jets to reduce gluon final-state radiation dependence. 

ATLAS selects online events with one hadronic jet with $p_T>$380 GeV, increased after offline reconstruction to 440 GeV, and at least an additional jet with $p_T>$60 GeV. For most of the benchmark signal considered it's required that $|y^*|\equiv|y_{j_1} - y_{j_2}|/2<$0.6, except for the $W^*$ search which results in being more performant selecting jet pairs with $|y^*|<$1.2. The aforementioned requirements allow for the search in a smooth spectrum in the range $m_{jj}>$1.1 TeV ($m_{jj}>$1.7 TeV for the $W^*$ search). 

Since the Monte Carlo (MC) simulations are not able to reproduce the QCD background $m_{jj}$ distribution, such smoothly falling background is fitted by CMS in its entire range, up to 8 TeV, with an analytical function $dN/dm_{jj} = p_1\cdot (1-z)^{p_2}\cdot z^{p_3}\cdot z^{p_4\ln z}$, where $z=m_{jj}/\sqrt{s}$. The variable number of parameters $p_i$ needed is driven by the fit complexity and usually increase as the collected luminosity increase.

\begin{figure}[!htb]
\centering
\subfigure[]
{\includegraphics[width=0.49\textwidth]{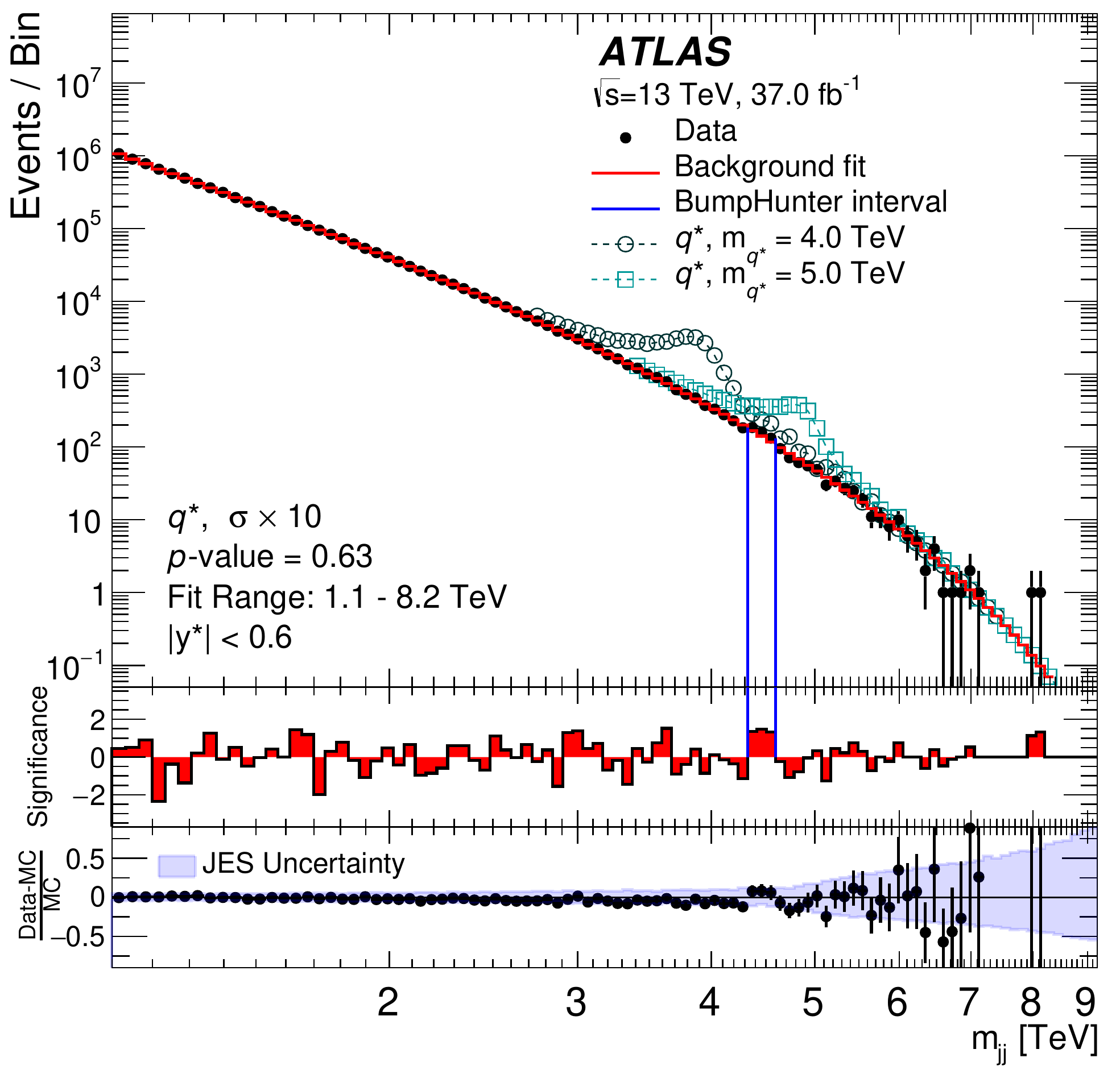}}
\hspace{0mm}
\subfigure[]
{\includegraphics[width=0.49\textwidth]{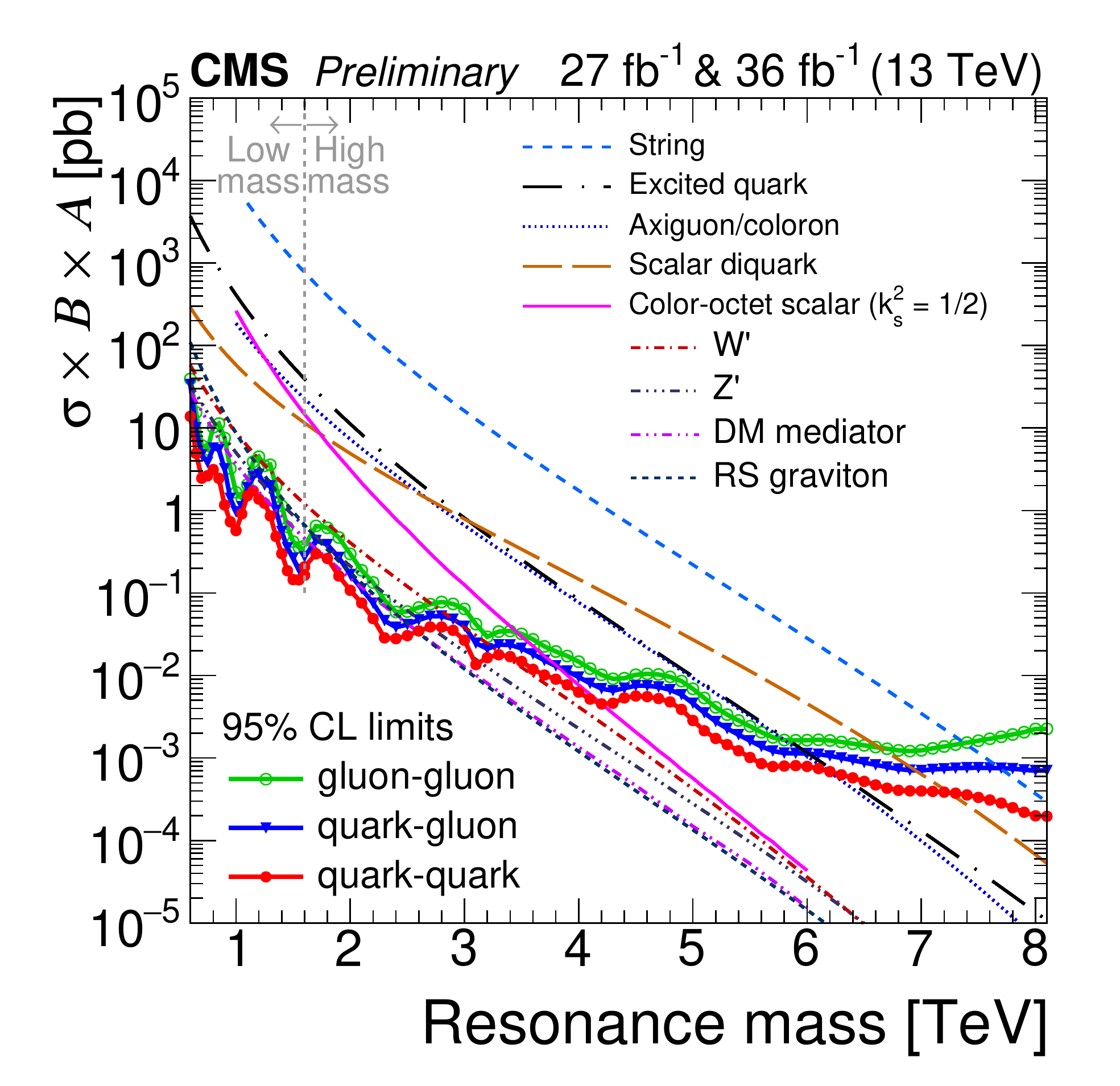}}
\caption{(a) Dijet invariant mass distribution \cite{atlas:dijet} observed by ATLAS in 37 fb$^{-1}$ of data collected (dots) compared to the background only prediction (red solid line); the lower panels show the residual significance and the jet energy scale uncertainty relevance. (b) The CMS limits \cite{cms:dijetRes} on generic signals for the $qq$, $qg$, and $gg$ combinations, compared to the theory predictions for the considered benchmark signals.}
\label{fig:figure1}
\end{figure}

To avoid possible fit instability ATLAS developed a background prediction technique based on a Sliding Window Fit (SWiFt): a fit in the restricted range (\emph{window)} of the spectrum with a simplified version of the analytical function used by CMS ($p_4$=0) is used to get QCD background prediction in the window center. Sliding the considered window through the spectrum it is possible to get a background prediction which is more stable against increase in the data sample statistics. Figure \ref{fig:figure1}(a) shows the $m_{jj}$ spectrum observed by ATLAS, together with the deviation significance of each bin with respect to the background prediction.

Neither ATLAS nor CMS found any significant deviation in the $m_{jj}$ spectrum with respect to the QCD background only hypotesis hence extracted limits on several models, simulated using different MC generators (details in \cite{atlas:dijet}, \cite{cms:dijetRes}). As a reference, the presence of excited quarks has been excluded up to particle masses of 6.0 TeV by both experiments, while additional $W^\prime$ bosons are excluded up to 3.6 TeV and 3.3 TeV by ATLAS and CMS respectively. These results has been provided also for a general reinterpretation as limits on the production cross section of more generic signals. Through the identification of $q$ and $g$ nature of the reconstructed jets CMS provided different limit result depending on the three possible combination of $q$ and $g$ in the final state. These limits are summarized in Figure \ref{fig:figure1}(b). ATLAS factorized out the detector effect on jet energy determination and provided limits for generic gaussian signal of a given width at particle level \cite{atlas:dijet}.

\section{Dijet angular searches}
\label{sec:angular}
Possible SM extensions might be characterized by an energy scale beyond the ones reachable by the current LHC collisions, approximately few tens of TeV. These won't therefore appear directly as new resonances but their low-energy range effect can be noticeable as a modification of the dijet event topology. An interesting benchmark model to consider is the quark-quark modified Contact Interaction (CI) \cite{contactInteraction}, which modifies the usual $q\overline{q}\to q\overline{q}$ scattering at a typical scale $\Lambda$. The effect of such CI is expected to modify the dijet angular correlation spectrum, namely the variable $\chi=e^{2|y^*|}$, also at $m_{jj}$ ranges within ATLAS and CMS current reach. The differential cross section for SM QCD quark scattering is approximately flat in $\chi$ for dijet events, while on the other hand CI-like events are expected to have more central jets, concentrating towards low $\chi$ values. 

\begin{figure}[htb]
\centering
\subfigure[]
{\includegraphics[width=.55\textwidth]{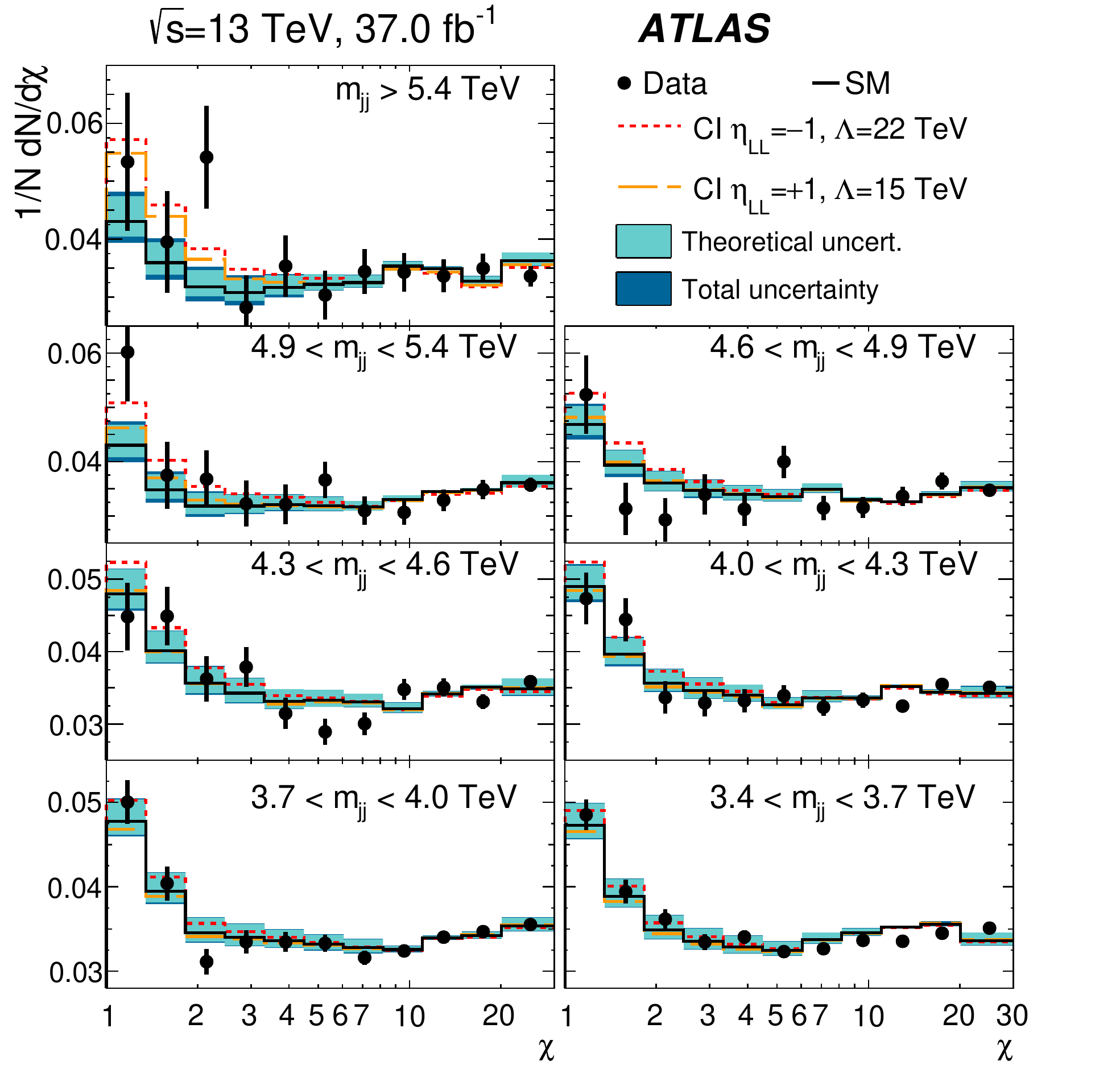}}
\hspace{0mm}
{\includegraphics[width=.43\textwidth]{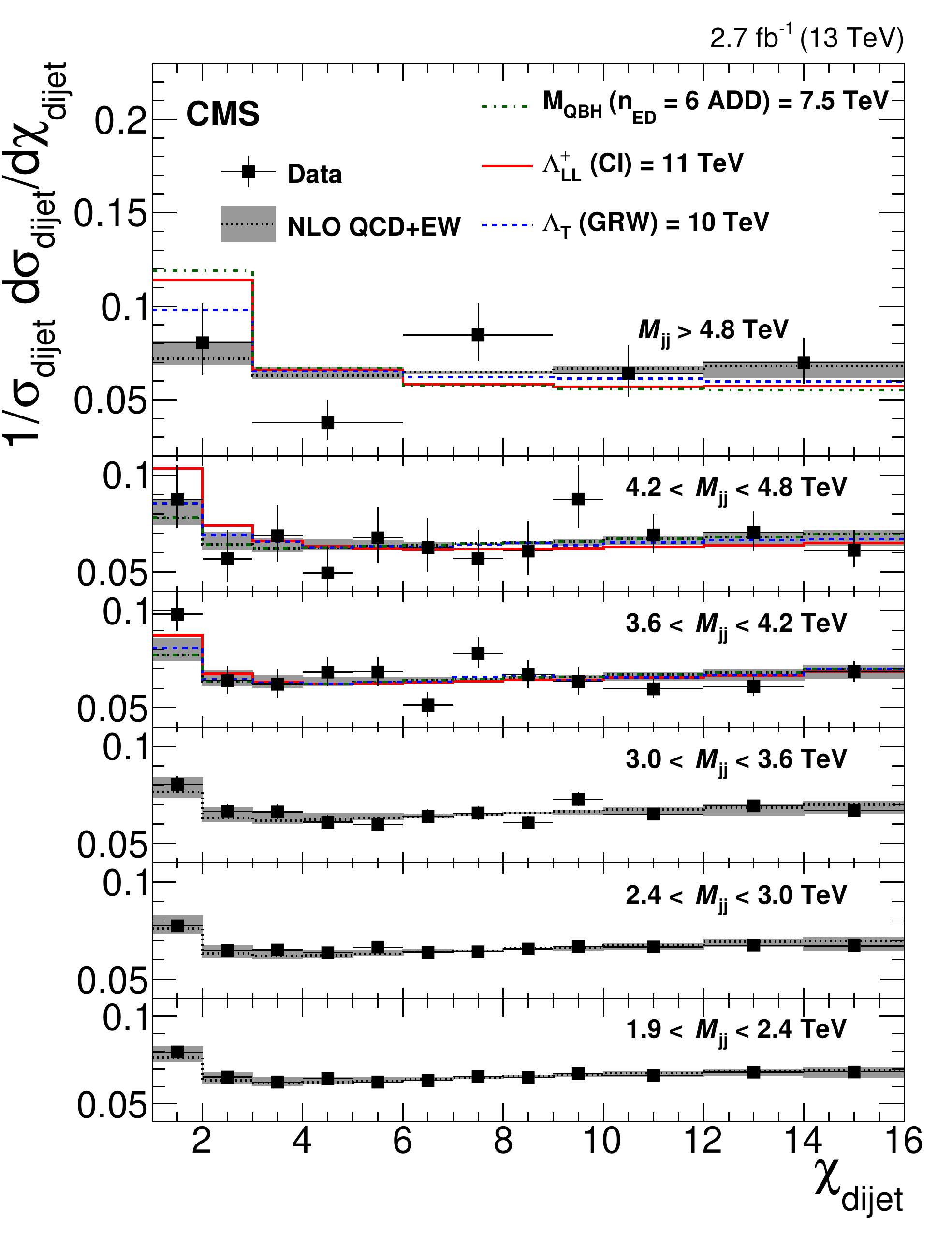}}
\caption{(a) The $dN/d\chi$ distribution of dijet events, for the different $m_{jj}$ range considered, and a comparison with SM background and possible CI signals as collected by (a) ATLAS \cite{atlas:dijet} and (b) CMS \cite{cms:dijetAng} experiment. }
\label{fig:figure2}
\end{figure}

The dijet event sample selected by ATLAS \cite{atlas:dijet} and CMS \cite{cms:dijetAng} for this kind of signature is similar to the one for a resonant signal search described before, but with an extended range in $y^*$, up to 1.7 and 1.39 for ATLAS and CMS respectively. To preserve the smoothness of the expected background the explorable mass spectrum is restricted to $m_{jj}>$1.9 TeV and 2.5 TeV by CMS and ATLAS respectively. The $dN/d\chi$ spectrum in the data, in different $m_{jj}$ bins, is compared to MC simulation prediction including most precise Next-to-Leading Order QCD corrections and Leading Order Electroweak effects. This spectrum is shown in Figure \ref{fig:figure2}(a) for the full dataset collected by ATLAS, including the overlaid prediction for CI signals with different $\Lambda$ value hypotheses. 
CMS looked in a similar way to the dijet angular distribution using the 2.7 fb$^{-1}$ of data collected during 2015, correcting the event kinematic to the particle level through a jet energy unfolding procedure, the result of which is shown in Figure \ref{fig:figure2}(b).

Since no significant deviation has been noticed with respect to the background only hypotesis a combined fit of the data to the MC prediction in the different $m_{jj}$ bins has been deployed to set lower limits on the characteristic scale $\Lambda$ of different exotic signal benchmark models. ATLAS excluded the presence or left-left (LL) chiral CI contribution up to 21.8 TeV assuming a constructive interference with the SM QCD, while excluded the range of $\Lambda$ below 13.1 TeV and between 17.4 and 29.5 TeV in the case a destructive interference is assumed. CMS provided limits in many different version of CI models (details in \cite{cms:dijetAng}) and in particular up to 14.7 and 11.5 TeV for the LL destructive and constructive interference model respectively.  

\section{Low-mass exploration}

The lower bound of the investigated dijet mass range for the searches described in the previous sections are driven by the requirement of online selection system (trigger) on jet kinematic, being the leading jet $p_T$ for the ATLAS case or $H_T$ for the CMS one. A lower bound in one of such variable, combined with the dijet event topology, imposes a restriction on the considered $m_{jj}$ range if a smooth background behaviour is desired, to avoiding turn-on effects. The minimum $p_T$ required for a trigger algorith is driven by the maximum affordable rate of events that can be recorded.
To overcome this limitation two ways have been deployed by the two collaborations: analyze events partially reconstructed \emph{online} and consider dijet events produced in association with a gluon or a photon from initial state radiation (ISR), which is reducing the overall rate of such dijet events.

\subsection{Trigger level data analysis}

\begin{figure}[htb]
\centering
\subfigure[]
{\includegraphics[width=0.46\textwidth]{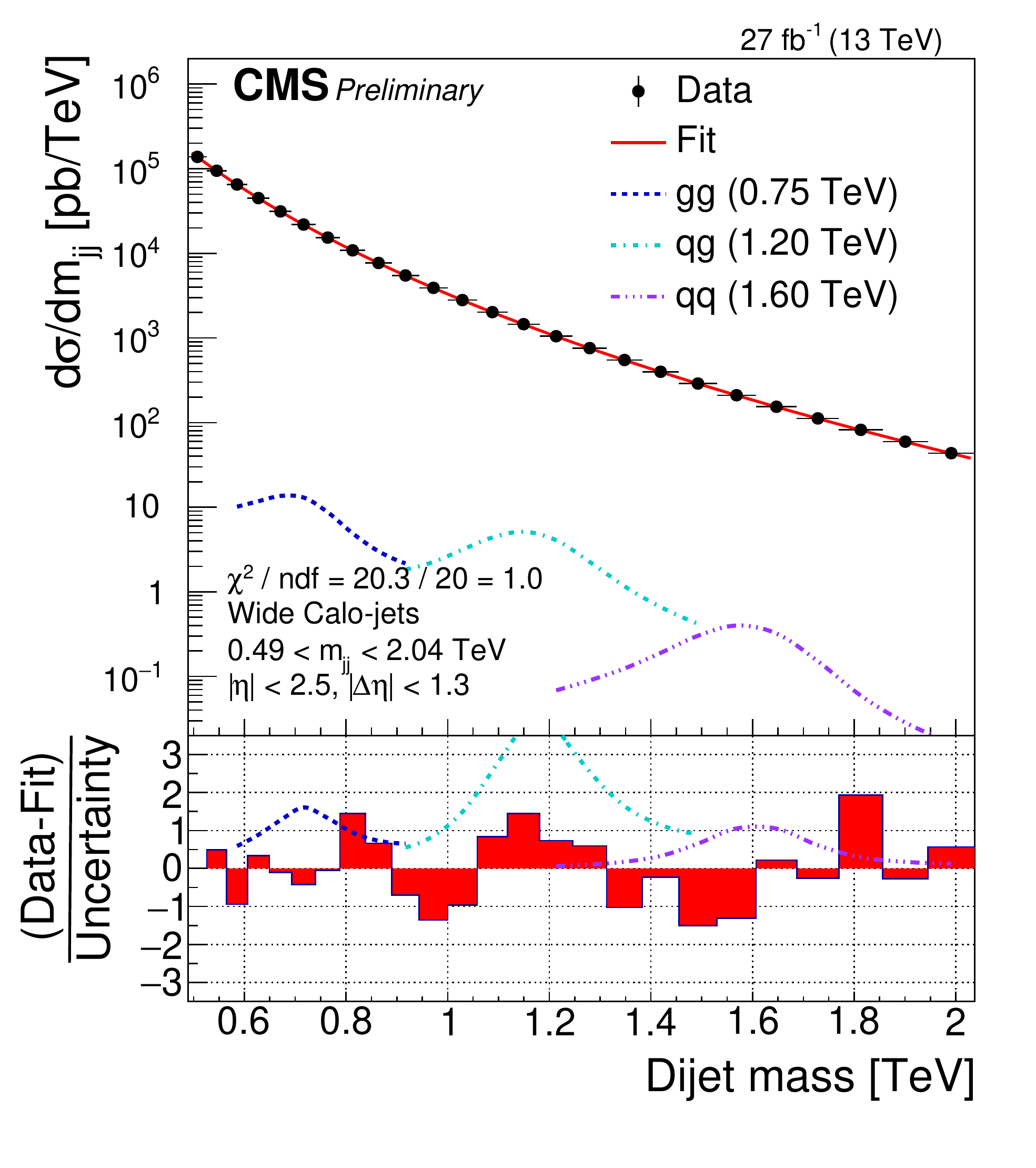}}
\subfigure[]
{\includegraphics[width=0.52\textwidth]{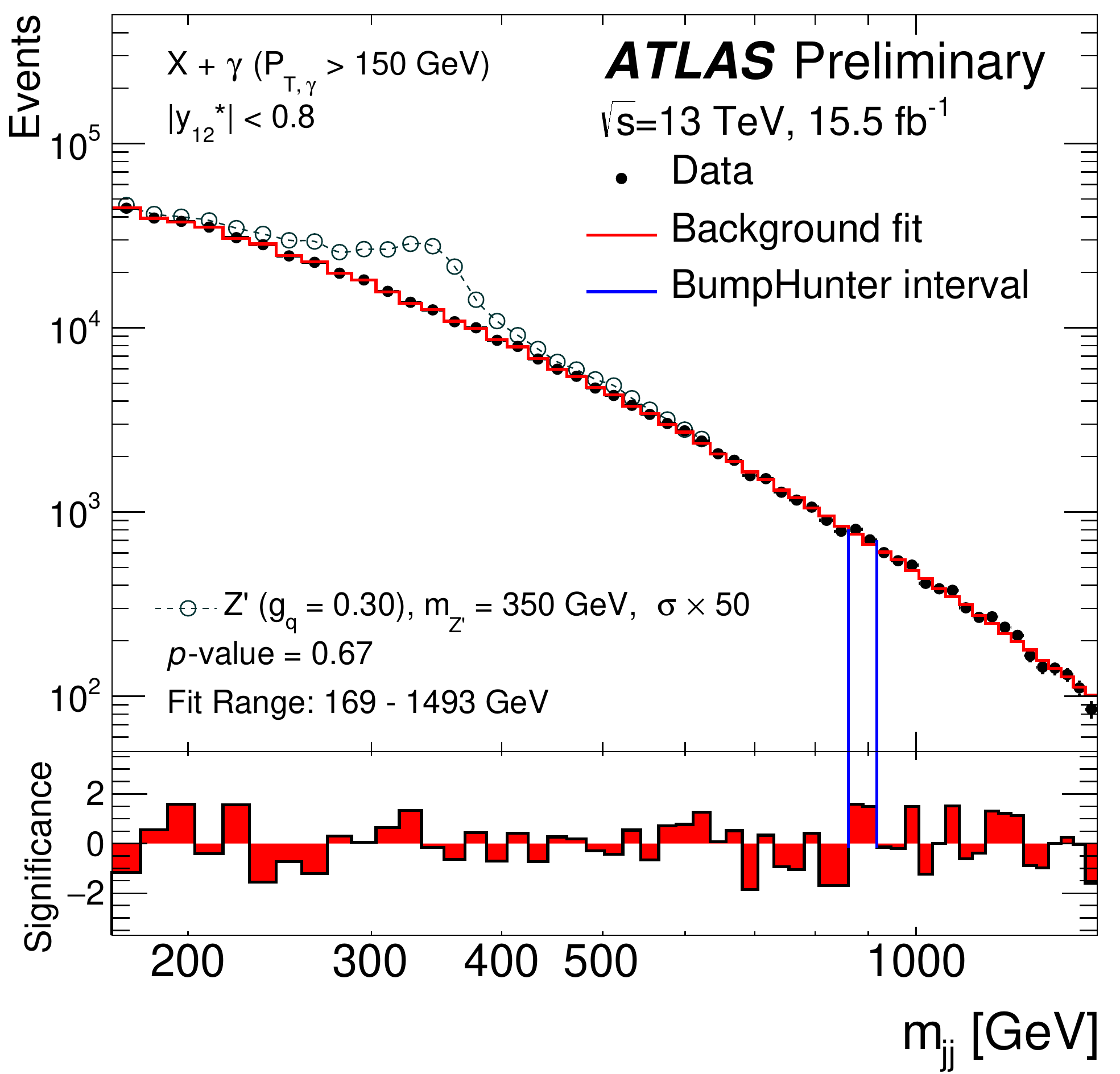}}
\caption{(a) The dijet invariant mass spectrum \cite{cms:dijetRes} for events recorded by CMS with the online data scouting reconstruction: black dots represent data while the solid red line is the result of the fit. (b) The invariant mass spectrum of dijet events produced in association with an ISR photon as reconstructed by ATLAS \cite{atlas:dijetisr}, compared to the result of the background-only fit.}
\label{fig:figure3}
\end{figure}

A jet-based trigger algorithm has a $p_T$ threshold driven by the data throughput that can be stored permanently, which depends itself on the rate of the (inclusive) process considered times the size of the event to be processed and recorded. Both ATLAS and CMS developed algorithms to process events from a detector partial reconstruction: processing only the calorimeter information allows for higher acceptance rate and reduced jet $p_T$ threshold. In the latest analysis CMS \cite{cms:dijetRes} reconstructed from high-level trigger calorimeter information jets with $p_T>$40 GeV and $|\eta|<$2.5 and recorded those with $H_T>$250 GeV and $|\eta_{j_1}-\eta_{j_2}|<$1.3. These selection criteria allow to collect, at a rate up to 4 KHz, a large sample of events to analyze with a dijet invariant mass in the range [0.49--2.0] TeV. The $m_{jj}$ spectrum, which is shown in Figure \ref{fig:figure3}(a), is fitted with an analytical function in a search for anomalous bump due to new physic resonances with the same strategy as the one discussed in Section \ref{sec:resonant}. 
The lack of anomalies has been interpreted as exclusion limits on production cross section for the benchmark models described in Section \ref{sec:resonant}, and shown in the low mass side (left) of Figure \ref{fig:figure1}(b). Signal with production cross sections greater than 0.1-10 pb have been excluded by CMS in the $m_{jj}$ range [0.49--2.0] TeV, which translate into the exclusion of the existence of any resonance from the considered signal models in the analyzed mass range, but for an RS-graviton, the existence of which is excluded up to 1.7 TeV. 

A similar analysis has been carried out by ATLAS, though using only about one tenth of the data collected by CMS, therefore providing less stringent constraint \cite{atlas:tla}.

\subsection{Dijet event associated to ISR jet or photon}

A different strategy to reach lower dijet invariant masses is searching for events with an additional gluon ($g$) or photon ($\gamma$) produced in the collision ISR. Such high-energy jet or photon can be the object satisfying the trigger requirements (see for example those mentioned in Section \ref{sec:resonant}), allowing to reconstruct the additional softer two jets potentially coming from the new physics resonance.

The preliminary result from ATLAS \cite{atlas:dijetisr} analyses 15.5 fb$^{-1}$ of data, selecting events with an ISR jet with $p_T>$440 GeV or a photon with $p_T>$150 GeV, with at least an additional pair of jets with $p_T>$25 GeV and $|y_{j_1}-y_{j_2}|/2<$0.6 or 0.8 respectively (for the $g$ or $\gamma$ ISR).
The $m_{jj}$ spectra in the range between [303-611] GeV and [169-1493] GeV, for the $g$ and $\gamma$ ISR case respectively, have been fitted with a smooth function, but no evidence of new physics has been found. Figure \ref{fig:figure3}(b) shows such spectrum for the $\gamma$ ISR dijet events.
In the two aforementioned signatures generic gaussian signals have been excluded in the sought mass ranges with cross sections greater than $\sim$0.1--0.01 pb depending on the mass and width of the resonance itself.

\begin{figure}[htb]
\centering
\subfigure[]
{\includegraphics[width=0.48\textwidth]{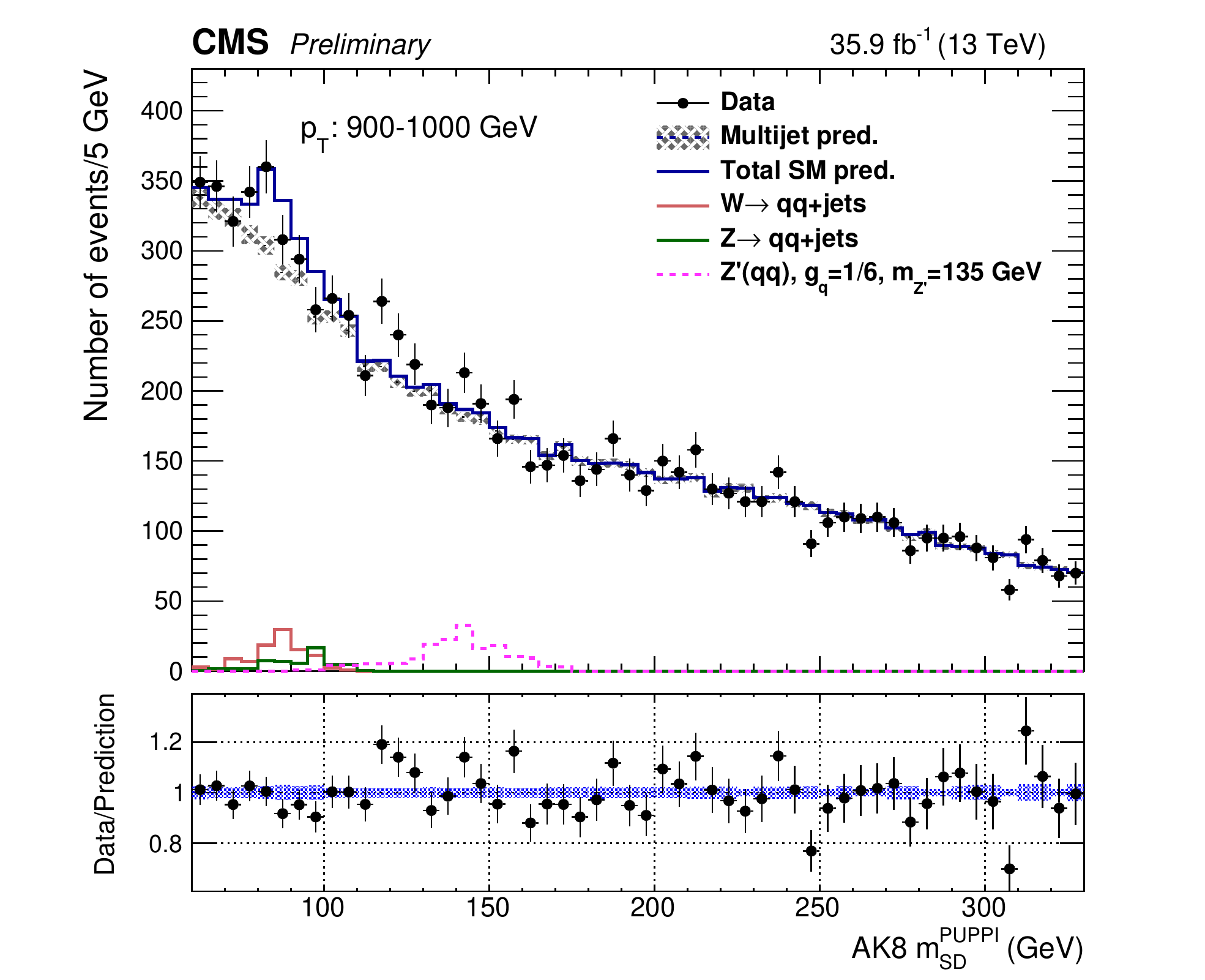}}
\subfigure[]
{\includegraphics[width=0.50\textwidth]{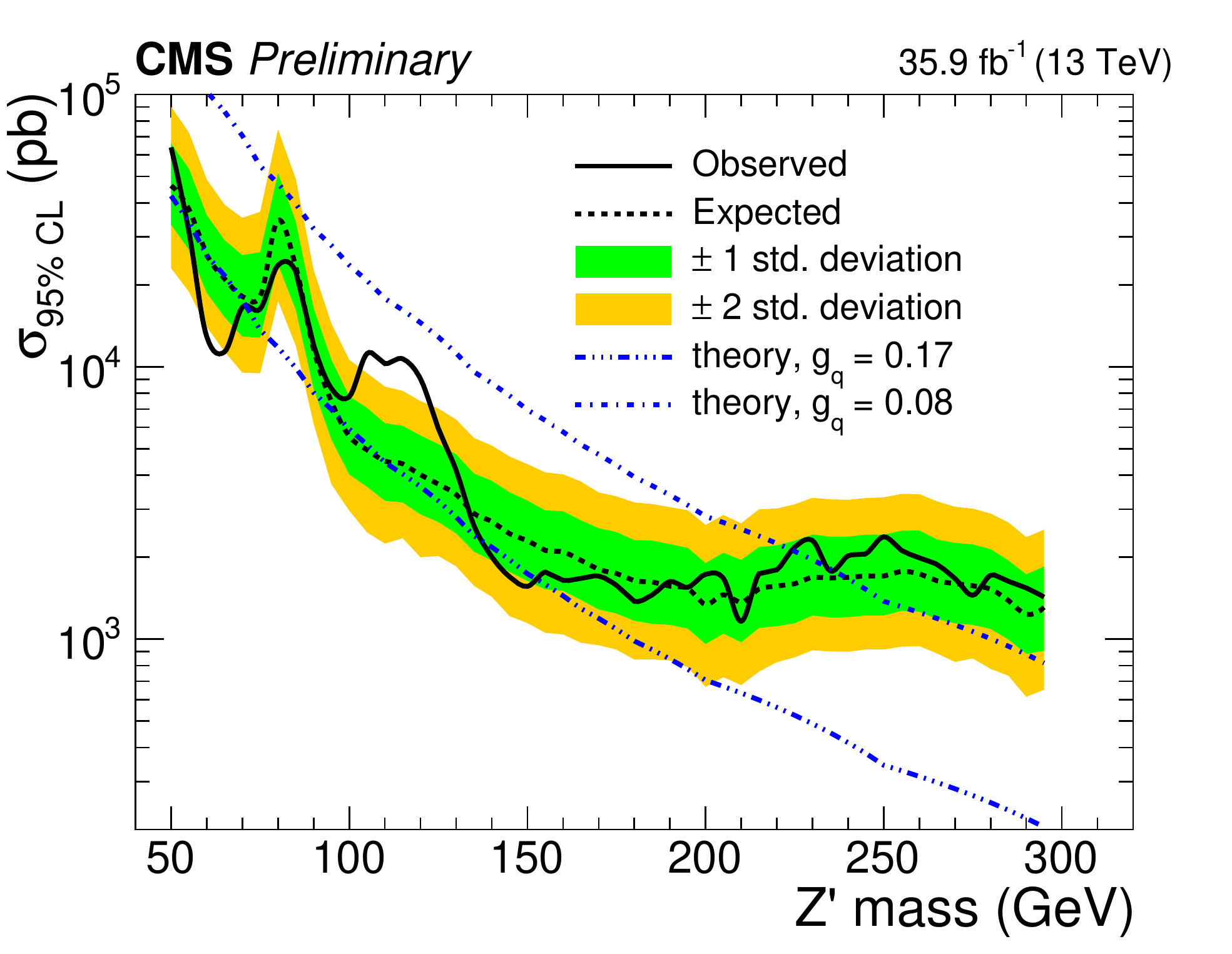}}
\caption{(a) Soft drop \cite{softdrop} jet mass distribution for the highest $p_T$ considered in the fit by CMS. (b) CMS 95\% CL upper limits on the $Z^\prime$ production cross section compared to the theoretical cross section \cite{lhc:dm}.}
\label{fig:figure4}
\end{figure}

A very low mass resonance (as it could be the SM $Z$ or $W$ boson) recoiling against and high energy ISR jet results in being boosted with a transverse momentum of the order of the ISR jet. In this kinematic regime the two jets produced in the resonance decay are so close to each other that they are often reconstructed as a sigle wide jet. To search for new low mass resonances CMS \cite{cms:isrwidejet} analyzed therefore a sample of events with at least a narrow jet reconstructed with the particle flow algorithm and $p_T>$500 GeV, $|\eta|<$2.5. The resonance decay system is reconstructed as a single wide jet to which jet substructure algorithms have been applied to identify its specificity \cite{substructure,softdrop}. The wide-jet mass spectrum has been analyzed in different bins of wide-jet $p_T$, comparing the data distribution to a multijet background prediction obtained from data control regions and the minor ones from $V\to q\overline{q}$+ jets and $t\overline{t}$ obtained using MC simulations. This comparison highlights an excess of events in the data, with respect to the background prediction, clustered at a mass of 115 GeV, for high values of resonance $p_T$, as can be seen in Figure \ref{fig:figure4}(a). The reported excess has a local p-value of 2.9 $\sigma$ and a global p-value of 2.2 $\sigma$. Limits have been set on the production cross section of a $Z^\prime$-like signal in the mass range [50,300] GeV, shown in Figure \ref{fig:figure4}(b), accessing for the first time the region below $\sim$100 GeV.

%

\section{Concusions}

Both ATLAS and CMS carried out general searches for resonances in hadronic dijet final states, with different dedicated strategy to explore, overall, a mass range spanning from 50 GeV to few TeV. Other than the specific models considered, the results obtained are interesting to be interpreted in the context of simplified Dark Matter models, for which the dijet signature is complementary to mono-object ones \cite{lhc:dm}.





\end{document}